# Decoupling electrocaloric effect from Joule heating in a solid state cooling device

M. Quintero[a,b], F. Gomez-Marlasca[a], L. Ghivelder[c], and F. Parisi[a,b].

a *Departamento Materia Condensada, CAC, CNEA, Av. Gral Paz 1429, San Martin (1650), Argentina*

b *Escuela de Ciencia y Tecnología, UNSAM, San Martín (1650), Argentina*

c *Instituto de Física, Universidade Federal do Rio de Janeiro, Cidade Universitária, Rio de Janeiro, 21941-972, Brazil*

**Abstract**

We report a heat dynamics analysis of the electrocaloric effect (ECE) in commercial multilayer capacitors based on $BaTiO_3$ dielectric, a promising candidate for applications as a solid state cooling device. Direct measurements of the time evolution of the sample's temperature changes under different applied voltages allow us to decouple the contributions from Joule heating and from the ECE. Heat balance equations were used to model the thermal coupling between different parts of the system. Fingerprints of Joule heating and the ECE could be resolved at different time scales. We argue that Joule heating and the thermal coupling of the device to the environment must be carefully taken in to account in future developments of refrigeration technologies employing the ECE.

The search for new materials for applications in solid state cooling devices is nowadays one of the most important topics in the field of alternative refrigeration methods.[1,2] The main idea behind solid state refrigeration is to produce changes in the temperature of a material as a result of the alignment of its magnetic or electric moment, which occurs on exposure to an external magnetic or electric field. The main advantage of these cooling techniques are an increase efficiency and minimized risk to the environment compared with the currently employed compressed gas systems. Solid state refrigeration technology requires the capability of the external magnetic or electric field to induce entropy changes near phase transitions on ferromagnetic or ferroelectric materials. The former phenomenon is known as the magnetocaloric effect[3] (MCE) and the latter as electrocaloric effect (ECE). In the simplest case, in both effects the external field induces a decrease of entropy due an increase of dipolar ordering.

The ECE was first measured on a Rochelle salt back in 1930.[4] Several decades later, when ferrolelectric materials with large polarization were discovered and developed, the interest on ECE was renewed.[5,6,7,8] The use of thin films[9,10,11] instead of bulk materials allows the possibility of applying high electric fields with relatively small voltages. Direct measurements of the ECE in bulk PbNb(Zr,Sn,Ti)O yield a maximum temperature variation of 2.5 K with 750 V (Ref. 12). On the other hand, on thin films of Zr-rich Pb(Zr,Ti)O (Ref. 13) a higher temperature change of 12 K was observed with a much smaller voltage of 25 V. The main disadvantage of thin films is the fact that they are thermally anchored by substrates, which reduces significantly heat pump capability and therefore the cooling power. A compromise solution is to use a multilayer film structure with metallic electrodes. Compared with films and bulk materials, the multilayer structures do not require a substrate and increase the volume of the electrocaloric material. This enhances the efficiency of the heat exchange yet keeping the high breakdown field of films. In a step forward toward actual applications, recent studies[14,15] measured the ECE in commercially produced multilayer capacitors (MLC). A temperature change of 0.5 K in an electric field E = 300 kV/cm was reported.[15] In spite of the rather weak magnitude of the effect it is clear that multilayer capacitors are promising candidates for applications as electric cooling devices, with an appropriated optimization of the geometry and material used.[16]

Cooling devices based on the magnetocaloric effect require large magnetic fields and expensive magnetic materials. On the other hand, electrocaloric devices require cheap ferroelectric materials, and large electric fields are easy to generate. However, an aspect so far neglected is the self heating of the electrocaloric device due to the Joule effect associated with drain currents. At first sight this effect may appear to be negligible when compared with the ECE contribution in specially optimized multilayer capacitors. But the possible large scale development of electrocaloric materials for cooling applications based on thermodynamic cycles demands a thorough evaluation of this supposedly small contribution, which is essential for a realistic evaluation of the cooling efficiency of the devices. In this Letter we present direct measurements of ECE in multilayer capacitors. An analysis of the heat balance equations allows us to decouple the temperature changes due to Joule heating and due to ECE. The different time scales inherent to each effect emerges as an important result, which must be considered in applications of the ECE in solid state refrigeration.

The sample investigated is a commercially manufactured MLC,[17] with a Y5V dielectric based on doped $BaTiO_3$ and nickel electrodes. A platinum thermometer (Pt-1000) was attached on top of the MLC using vacuum grease to ensure a good thermal contact. The MLC is then placed on a Cu block using 2 mm Teflon layer for thermal isolation. The block is placed in a vacuum chamber. The voltage was applied using a Kethley-2400 source and an Agilent 34401 multimeter was used to measure both the drain current and the thermometer resistance. All measurements were performed at room temperature. A schematic view of the experimental setup is shown in Fig. 1a.

In Fig. 2 we present results of the excess temperature registered by the thermometer $\Delta T_M = T_H - T_0$ (where $T_H$ and $T_0$ are the temperatures of the sample holder and thermal bath respectively) as a function of time during the application of different voltages on the MLC. After the voltage is applied a sudden increase in $\Delta T_M$ is observed due to the ECE, followed by a slow relaxation associated with the coupling to the thermal bath. At higher voltages the temperature reached after relaxation is larger than the starting temperature due to the effect of Joule heating in the MLC. An abrupt decrease of $\Delta T_M$ is observed when the voltage is turned off, an ECE with decreasing voltage, followed by a slow relaxation to the base temperature $T_0$ (the Joule term is absent).

To describe quantitatively the temperature change we model the system following the schematic diagram shown in Fig. 1b. Within this framework, the set of heat balance equations to be satisfied are.[18]

$$\frac{dQ_S}{dt} = \gamma_{SH}(T - T_H) + \gamma_E^S(T - T_0) = -C_S \frac{dT}{dt} + iV + \xi(E)\frac{dE}{dt} \quad (1)$$

$$\frac{dQ_H}{dt} = \gamma_{SH}(T_H - T) + \gamma_E^H(T_H - T_0) = -C_H \frac{dT_H}{dt} + I_{Pt}V_{Pt} \quad (2)$$

Equation (1) describes the heat exchange by the sample with the source of heat taken as the ECE and the Joule heating introduced by the drain currents. Equation (2) describes the sample holder, and contemplates the Joule effect occurring in the thermometer. T is the temperature of the sample and $T_H$ that of the sample holder; the latter is the actual temperature measured by the Pt-1000 thermometer. $T_0$ is the temperature of the thermal bath, E is the electric field over the Y5V dielectric, $\gamma_E^S$ and $\gamma_E^H$ are the thermal conductance between the thermal bath and the sample and sample holder respectively, and $C_S$ and $C_H$ are the heat capacities of the sample and the sample holder respectively. $\gamma_{SH}$ is the thermal conductance between the sample and the sample holder, and $I_{pt}$ and $V_{pt}$ are the current and voltage over the Pt-1000 thermometer. $I_{Pt} = 0.01$ mA was used in the whole experiment in order to neglect this term. The experimental setup was designed to ensure a good thermal contact between the sample and the sample holder, and the measurements were performed under high vacuum conditions, so it is expected that $\gamma_E^S \cong \gamma_E^H \cong \gamma_E << \gamma_{SH}$. To obtain the value of $C_H$ we used the adiabatic heat pulse calorimetric technique,[19,20,21] with the Pt-1000 resistance acting both as thermometer and heater, and without any voltage applied to the sample. We found $C_H = 0.15$ J K$^{-1}$. For obtaining $C_S$ we use the calculated heat capacity of the MLC, $C_S = 0.04$ J K$^{-1}$ (Ref. 14).

Qualitatively, three different processes can be distinguished with very different characteristic time scales. Initially we suppose a change in the sample temperature due to the ECE, with a characteristic time $\tau_s \approx 20ms$ associated to the time taken by the power source to reach its final voltage. For $t \leq \tau_s$ it is assumed that the sample does not exchange heat with the other components of the system while the electric field is raised, and therefore the Joule term can be neglected within this time interval. Subsequently, the sample reaches a temperature $T_0 + \Delta T_1 = T_0 + \frac{1}{C_S}\int_0^{E\max}\xi(E)dE$ immediately after the electric field is applied. The following process ($t \leq \tau_{SH}$) is the stabilization of the temperature of the sample with the Cu base, still neglecting the coupling with the environment, and considering the Joule term as the only source of heat. This process is registered by the thermometer in characteristic times of a few seconds, $\tau_{SH} \approx \frac{C_H C_S}{C_H + C_S}\gamma_{SH}^{-1}$, representative of the thermal coupling between the sample and the sample holder in which the thermometer is placed. Finally, a temperature relaxation is observed in times of the order of tens of seconds, associated with the heat exchange between the sample and the sample holder with the thermal bath. The characteristic time related to this process can be estimated as $\tau_E \approx (C_H + C_S)\gamma_E^{-1}$. The experimental setup has been designed to obtain a good thermal contact between the sample and the sample holder, and the measurements were performed under high vacuum conditions to ensure thermal insulation with the thermal bath, so it is expected that $\tau_{SH} \ll \tau_E$.

The entire process described by Eq. (1) and (2) is equivalent to a couple of over-damped oscillators. The time dependence of the temperature registered by the thermometer after the electric field is applied is then described by the sum of two independent exponential decays

$$T_H = T_0 + T^* + A_1 e^{-\frac{t}{\tau_{SH}}} + A_2 e^{-\frac{t}{\tau_E}} \qquad (3)$$

The parameters obtained by fitting the experimental results with the Eq. (3) are shown in Fig. 3, both when the field is turned on and turned off. As expected, very different values for $\tau_{SH}$ and $\tau_E$ are obtained: $\tau_{SH}$ = 2.05 ± 0.26 sec. and $\tau_E$ = 23 ± 9 sec (Fig. 3a). The large error observed in $\tau_E$ could be related with the raw approximation of the model in the intermediate timescale (between $\tau_{SH}$ an $\tau_E$).

More interestingly, the parameter $A_1$, which describes the process at times of the order of $\tau_{SH}$, depends only on the magnitude of the voltage change, and not on its sign, indicating that the Joule effect is not relevant at a time scale determined by $\tau_{SH}$ (Fig 3b). This fact allows us to obtain the value of the maximum temperature reached by the sample in adiabatic conditions as $\Delta T_1 = -\frac{C_H}{C_S} A_1$. This important result is a direct measurement of the real magnitude of the electrocaloric effect in the studied device. The temperature change obtained for 200 V (300 kV/cm) equals to 0.32 K, close to the calculated from indirect measurements using the Maxwell relation[14]. Finally, the parameter $T^*$ related with the stationary temperature is determined by the Joule term, $T^* \approx IV/2\gamma_E$. The behavior of this term is displayed in Fig. 3c, showing its dependence with the applied voltage, both before and after the voltage application. The linear relation between $T^*$ and $IV$ is shown in the inset of Fig. 3c.

In summary we presented a detailed study of the heat dynamics in a commercial multilayer capacitor. We argue that the relevance of Joule heating is a key factor for the development of electrocaloric-based cooling devices. We performed direct measurements of the temperature changes of the sample produced by the application of an electric field. To separate the temperature changes produced by the ECE and Joule heating it was necessary to model the heat exchange between the different elements in order to evaluate quantitatively and compare each contribution. The solution of the resulting equations corresponds to that of a set of coupled over-damped oscillators, with two well defined time scales, mainly determined by the thermal coupling between the components of the system. Substantial difference in these characteristics time scales allowed the effects of the ECE and Joule heating to be decoupled. This procedure allows us to determine through direct measurements the true magnitude of the ECE in this device.

Our study was conducted on a material (Y5V) selected to minimize the presence of leakage currents. It is expected that the performance of existing MLCs can be greatly improved via materials selection and geometric modification,[16] but for such enhancements to succeed the contribution of the Joule effect must be taken into account. Another aspect that plays an important role for application efficiency is the characteristic time for heat exchange with the thermal bath. We have shown that this is a key factor for developing future solid state cooling devices based on the ECE. On this aspect, the MLC ability to quickly conduct heat from the dielectric to the exterior appears to be one of its most important advantages.

We thank to N. Mathur for fruitful discussion and providing the studied sample.
Support from ANPCyT PICT 08-01327, CONICET PIP-11220090100889 and UNSAM is acknowledged. M. Quintero is a member of CIC CONICET. L. Ghivelder acknowledge financial support from FAPERJ and CNPq.

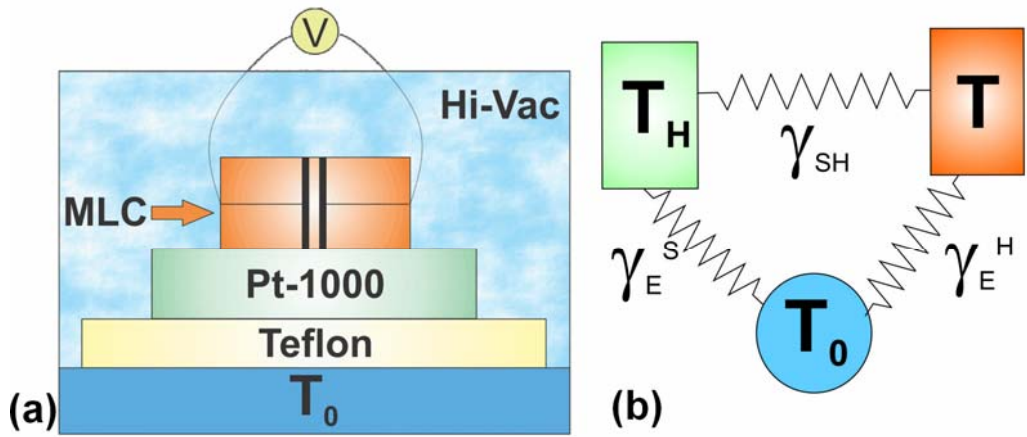

Figure 1: a) Schematic description of the experiment. b) heat exchange diagram for modeling the system. The sample and the sample-holder are connected with a thermal link $\gamma_{SH}$ gamaSH. At the same time, both parts are connected with a thermal bath at a constant temperature T0 through the thermal links $\gamma_E^S$ and $\gamma_E^H$ for the sample and the sample-holder respectively.

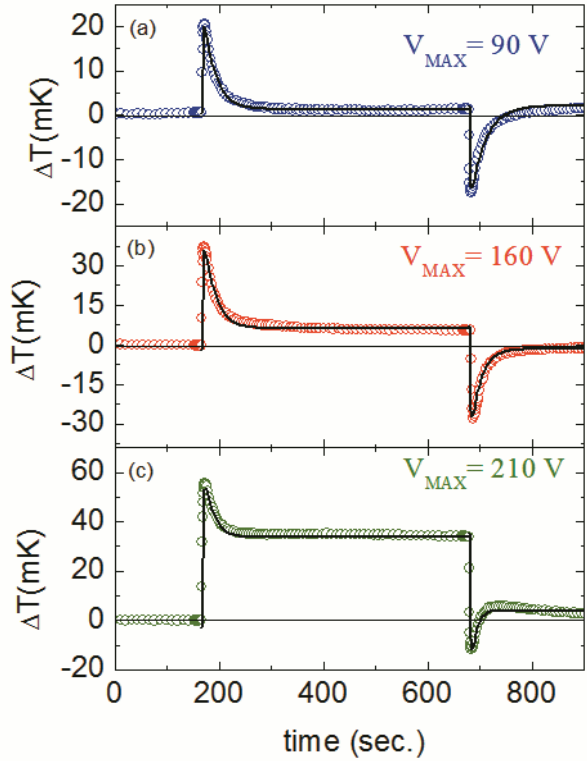

Figure 2: temperature difference as function of time for different applied voltages. The voltage is applied during 600 sec and then it is turned off. The lines are the fitting function using Eq. (3). Note the different scales on the *y* axis of the different panels.

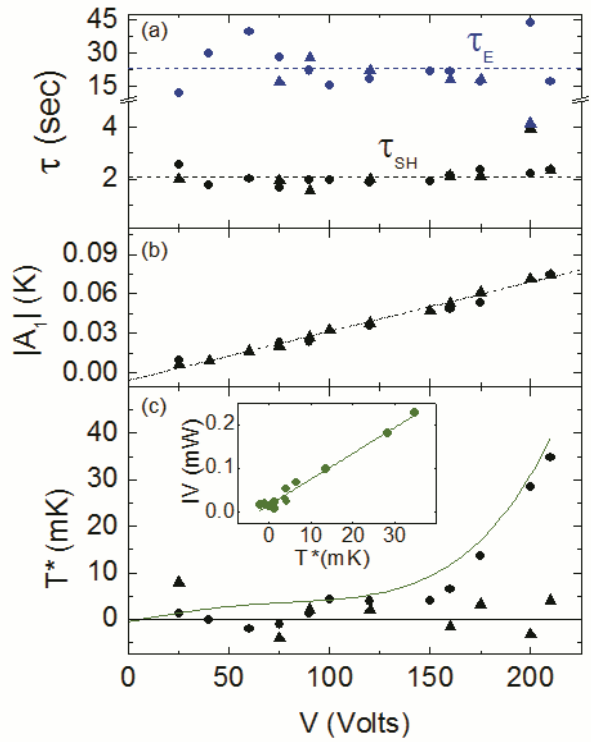

Figure 3: fitting parameters obtained from the measurements shown in Fig 2 at different applied voltages. Characteristic times (panel a), $A_1$ (panel b) and $T^*$ (panel c). The Joule

heating contribution is presented as a dotted line in panel (c). Circles and triangles indicate increasing and decreasing voltage, respectively.